# Initial conditions in the neural field model

Pedro A. Valdés-Hernández, Thomas Knösche

*Abstract*— in spite of the large amount of existing neural models in the literature, there is a lack of a systematic review of the possible effect of choosing different initial conditions on the dynamic evolution of neural systems. In this short review we intend to give insights into this topic by discussing some published examples. First, we briefly introduce the different ingredients of a neural dynamical model. Secondly, we introduce some concepts used to describe the dynamic behavior of neural models, namely phase space and its portraits, time series, spectra, multistability and bifurcations. We end with an analysis of the irreversibility of processes and its implications on the functioning of normal and pathological brains.

## I. INTRODUCTION

The way the dynamics of the brain is modeled is manifold (Deco et al. 2008; Coombes 2009). Neural dynamical models can target single cells, specific areas of the brain, e.g. the visual cortex or the patch covered by a grid measuring local field potentials (Pinotsis, Moran, and Friston 2012), or the whole neocortex (Robinson 1998); even including the thalamus and other subcortical nuclei (Rennie, Robinson, and Wright 2002). Neural modeling spans from assemblies of conductance-based or integrate-and-fire (IF) cells, to mean field approaches[1] such as *neural mass models* (NMM) and *neural field models* (NFM). The latter are based on the assumption that a *population* of a certain type of neuron (e.g. layer V excitatory pyramidal or basket inhibitory) can be represented by a single representative with average morphological and physiological parameters; being these lumped into a few set for the sake of simplicity and tractability (Deco et al. 2008).

The equations governing a dynamical model describe the evolution of the state variables of the system in time. Dependency of this evolution on initial conditions (namely $t = 0$) is relevant when *multistability* or *chaos* arises from the presence of nonlinearities in the equations. Setting different initial conditions can be the consequence of presetting different values of a certain subset of parameters of model, thus inducing *bifurcations* (Kuznetsov 1998); or due to external inputs non-parameterized by the model such as stochastic noise.

Despite the enormous amount of proposed models on the biophysical dynamics of the brain (Deco et al. 2008), we find that a systematic revision of the dependency of the dynamic evolution of the neural system on the *initial conditions* is lacking. In this review we collect some examples where multistability and bifurcations arises and discuss the implications of setting different initial conditions for the future evolution of the system. This might have an impact, for example, in the behavior of brains predisposed to epileptic seizures (Breakspear et al. 2006; Freyer et al. 2011; Spiegler et al. 2010), where abnormal changes in the value of parameters might induce "undesirable" bifurcations of the dynamical system toward irreversible high amplitude spiking.

In the second section of this paper we review different basics notions. Subsection A introduces some basic notions on NMM and NFT and show how they can be transformed to the canonical form of a dynamical system. Subsection B provides a brief introduction to the basic concepts on dynamical systems necessary to understand the temporal evolution of the solutions of neural models. The third section describes some examples of multistability and bifurcations, discusses how different initial conditions are fulfilled as well as their effects on the future of the system. The reversibility of the evolution of the systems is discussed.

## II. BASIC NOTIONS

### A. A General Overview of Mean Field Models

A mean field model describes the equations governing the evolution in time of the *membrane voltage* at the soma and *spike rates* of either a single or distributed neural masses, each comprising $P$ populations of neurons. Ubiquitous in all mean field models are the so-called pulse-to-wave and wave-to-pulse conversion (Jirsa and Haken 1997). The first describes how the incoming spike rate provokes changes in the membrane potential at the soma. Just for illustrative purposes we write down a general and very simplistic form of a neural field:

$$\begin{aligned} v_p &= \sum_{q=i,e} h_{pq} * \phi_{pq} + v_0 + \upsilon_p \\ \phi_{pq} &= P_{pq} \otimes Q_q \\ p,q &= 1 \ldots P \end{aligned} \quad (1.1)$$

In these equations $*$ and $\otimes$ temporal and spatiotemporal convolution operators respectively, $v_p$ is the membrane potential at the soma of population p, $v_0$ is a constant DC, $Q_q$ is the efferent spike rate of population q, $\phi_{pq}$ is the afferent spike rate propagating from the presynaptic population q to the postsynaptic population p and $\upsilon_p$ is the external input rate. The kernel $h_{pq}\left(\tau_{pq}^r, \tau_{pq}^d\right)$ is the postsynaptic response which is ruled by rise and decay times. The propagator or Greens function $P_{pq}$ encodes the

P. A. Valdés Hernández is with the Cuban Neuroscience Center, Havana, Cuba (e-mail: petermultivac@gmail.com).

T. Knösche is with the Max Planck Institute for Human Cognitive and Brain Sciences, Leipzig, Germany (e-mail: knoesche@cbs.mpg.de).

[1] through the Fokker-Planck equation

connectivity rules (characterized by certain spatial scales) and time delays (characterized by a distribution of propagation velocities) between neural masses (extrinsic) and between different populations within a mass (intrinsic). Intrinsic connectivity is almost always considered as isotropic and mostly shifts invariant[2], while extrinsic connectivities might be considered as isotropic or anisotropic. Isotropic connectivities allow (1.1) to be converted to a set of PDEs (Coombes et al. 2007). For a very peaked velocity distribution $c_0$:

$$v_p - v_0 = \sum_{q=i,e} v_{pq} + \phi_p$$

$$L\left(\frac{\partial}{\partial t}\right) v_{pq} = \phi_{pq} \qquad (1.2)$$

$$\frac{N\left(\frac{\partial}{\partial t}, \frac{\partial^2}{\partial t^2}, \ldots, \frac{\partial^n}{\partial t^n}, c_0^2 \nabla^2\right)}{D\left(\frac{\partial}{\partial t}, \frac{\partial^2}{\partial t^2}, \ldots, \frac{\partial^m}{\partial t^m}\right)} \phi_{pq} = Q_q$$

In this equation $L$ is a polynomial of order zero, one or two[3] and $D$ and N are also polynomials with $m < n$ to satisfy the Kramers-Kronig causality constraint (Bohren 2010). The lowest DC spatial mode, i.e. $\nabla^2 = 0$ (or $c \to +\infty$ inside (1.1)), results in a set of ODEs of a single mass. If we establish (1.2) for N masses we obtain a typical NMM.

There are different variants for the types of populations. The most usual is to take excitatory (e) and inhibitory (i) types. Excitatory (inhibitory) populations induce depolarization (hyperpolarization) in the postsynaptic populations thus $h_{pe} > 0$ ($h_{pq} < 0$). Other models incorporate a third population of excitatory spiny stellate neurons in layer IV (e) of the neocortex, leaving population E for the pyramidal cells (Jansen and Rit 1995). Recently, the "canonical" circuit was proposed where pyramidal in layer II/III was distinguished from those in layer V/VI (Bastos et al. 2012). More detailed models including up to 19 types of populations are found (Binzegger, Douglas, and Martin 2009). Specific (s) and reticular (r) thalamic nuclei have been also included (Robinson 2005).

The wave-to-pulse conversion describes the how average depolarization in the soma modifies spike rates. Through this equations (1.1) and (1.2) can be closed by obtaining the mean field approximation of the IF behavior of neurons. That is, once the membrane potential of a neuron goes beyond certain threshold it fires an action potential. For example, for a Gaussian distribution of thresholds within the population, the mean behavior of the thresholds follows an error function. This function has a sigmoidal shape thus different sigmoid functions are proposed for the ease of mathematical solvability. We shall generically state it as $Q_p = \text{Sigmoid}(v_p)$. *Here is where the static nonlinearity of the model resides*.

### B. A General Overview on Dynamical Systems

Since the mathematical tools for the dynamical analysis of PDE is quite complicated and still in development, we shall refrain to the analysis of ODEs like (1.2). This dynamical equation can be easily reformulated in a canonical form as:

$$\dot{\mathbf{x}} = \mathbf{f}(\mathbf{x}, \mathbf{u}, \boldsymbol{\theta}) \qquad (1.3)$$

In this equation we've encapsulated all the *state variables* in a state vector $\mathbf{x}$. The state variables can be the *membrane potential* or any of the *spike rates* as well as their derivatives up to an order which is determined by the original form of (1.2). The domain of the state variables is called the *state space*, i.e. $\mathbf{x} \in \mathbb{X}$. The vector $\mathbf{u}$ represents the inputs which can be deterministic or stochastic, while $\boldsymbol{\theta}$ is the vector of parameters.

A point in the state space defines a *state*. A solution of (1.3) $\mathbf{x}(t; \mathbf{x}_0)$ for $t \in (-\infty, +\infty)$ given an initial state $\mathbf{x}_0$ is called an orbit. We are interested in orbits with causal evolution ($t \geq 0$). The most important orbits in the analysis of dynamical systems are *equilibria* (or fixed points for discrete time systems) and *cycles*, which are special cases of the so-called invariant subspaces of $\mathbb{X}$. Fig. 1 and 2 shows examples of equilibria and cycles in a 2D state space, respectively. Equilibria are constant over time while cycles are periodic orbits. *Limit cycles* are special types of cycles: a subspace of $\mathbb{X}$ can be defined where no other cycle exists (Kuznetsov 1998).

Equilibria and limit cycles can be *stable* or *unstable*. Stable equilibria or limit cycles are called *attractors*. Stability of an equilibrium is determined by calculating the eigenvalues of (1.3), after being linearized in the vicinity of the equilibrium. For continuous time systems the equilibrium is stable if the real part of all eigenvalues is negative. Otherwise is unstable. A saddle point is when not all eigenvalues have a positive real part. A stable equilibrium can be a *node* or a *focus*, for real only or pairs of complex conjugates eigenvalues respectively. Complex eigenvalues yields oscillatory behavior (Kuznetsov 1998).

| Eigenvalues | Phase portrait | Stability |
|---|---|---|
| | node | stable |
| | focus | |
| | saddle | unstable |
| | node | unstable |
| | focus | |

Figure 1. Types of equilibria and their stability conditions. (Taken from (Kuznetsov 1998))

---

[2] For an exception see (Pinotsis and Friston 2011)
[3] We are not mentioning conductance based synapsis with shunting currents in this simplified model.

The stability of a limit cycle is determined by analyzing stability of the fixed point of its corresponding Poincaré map. The latter is a discrete time dynamical system with one fixed point which is defined as the state of $\mathbb{X}$ belonging to both the limit cycle and a plane not tangential to the limit cycle (see Fig. 3). If the eigenvalues of the Poincaré map are within the unit disk the limit cycle is stable. Otherwise is unstable, being a saddle if not all eigenvalues are outside the unit disk (Kuznetsov 1998).

In summary, the basic notion about stable attractors is that any transient orbit in their vicinity tends to them as $t \to +\infty$ [4]. Fig. 2 depicts stable and unstable equilibria and limit cycles as well as the behavior of transient orbits in time.

The partitioning of $\mathbb{X}$ into orbits is named the *phase portrait* of the dynamical system. The phase portrait gives an idea of the number and types of asymptotic states of the system. The simplest systems, i.e. linear ones, have a single equilibrium, which might be stable. Nonlinearity is the responsible of the multistable nature of the steady states of the system (Kuznetsov 1998). *Trivial multistability* implies several possible stable attractors for a given set of parameters. The basin of attraction of a stable attractor is the set of all initial conditions $\mathbf{x}_0 \in \mathbb{X}$ from where orbits will asymptotically tend to it. Therefore, for neural models defined by (1.2), *the final state of the brain will depend on basin of attraction in which the system was at the initial time point* (see the leftmost phase portrait in Fig. 2 for an example).

A system, for certain values of its parameters, can have basins of attraction with orbits showing chaotic behavior. For very close initial points in the basin, final states are quite dissimilar. This orbital separation is quantified by Lyapunov Exponents, which are the rates at which orbits separate exponentially. The largest LLE correspond to the dimension of maximal rate and thus leads the behavior for $t \to +\infty$. "Chaoticity" is quantified by positive LLEs. Chaotic behavior is characterized by non-periodic and unpredictable orbits. If they go toward a set of points or cycles this is called a *chaotic attractor*. Chaos and trivial multistability can coexist as *generalized multistability*.

For the neural model we can construct a *time series* associated to the behavior of the magnitude characterizing the mean field, e.g. membrane potential. The time series of a limit cycle is a periodic signal, while for an equilibrium is a constant value across time. For transient states near the vicinity of equilibria, time series grows infinitely for unstable equilibria. For nodes and foci, time series decay exponentially to zero, the signal oscillating in the last case. However, a time series could fluctuate indefinitely around a stable equilibrium provided that it receives stochastic inputs. In fact, stochastic inputs can lead to jumps into more than one basin of attraction in a multistable system. *Neural models could dwell among a repertoire of different generalized stable states without preference* (Deco and Jirsa 2012; Freyer et al. 2011). Stochastic inputs are ubiquitous in the brain. Thus, neurons are never inactive. A stochastic wandering among an ensemble of stable states (or even saddle points) might constitute the so-called resting state activity of the brain. A certain subset of attractors might become of the preference of the system after some external specific stimulus.

A change in the values of certain subset of parameters of the model can induce *topological changes* in the multistable configuration of the system, i.e. stable states can appear or disappear, or existing stable states can turn unstable or vice versa. These are called *bifurcations*. Bifurcation can be *local* (respect to an equilibrium) if the analysis of the orbit within an infinitesimal vicinity is enough to detect the bifurcation, or *global* otherwise. Both the parametric portrait and the phase portrait form a bifurcation diagram. The plot of the phase portrait, projected to a certain relevant dimension where the state variable bifurcates its orbits, versus only one changing parameter (codim1 bifurcation) is the common type of bifurcation diagram shown in the papers we reviewed (Kuznetsov 1998). A bifurcation diagram is stratified according to the basins of attractions and the bifurcation

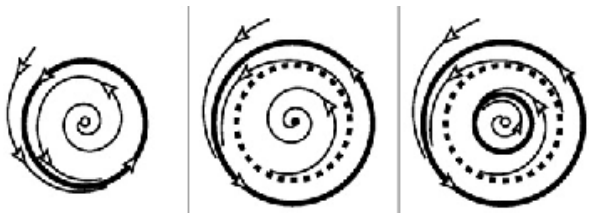

Figure 2. Phase portraits of limi cycles: stable (continous line) and unstable (dashed lines). Transient orbits are also represented. Note that they converge asymptotically to the limit cycle attractor depending on the basin of attractions in which they start. (Taken from (Spiegler et al. 2010)).

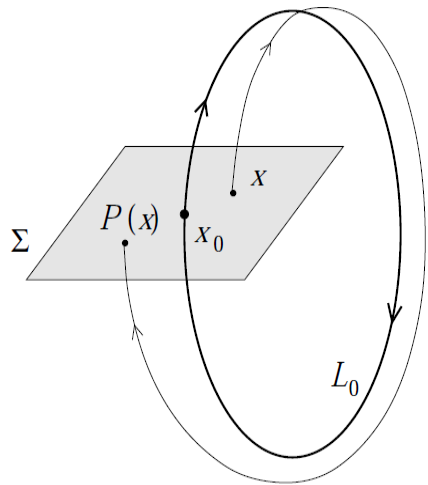

Figure 3. Graphical representation of a Poincaré map associated with the cycle $L_0$. The plane $\Sigma$ contains the fixed point $\mathbf{x}_0$ and different points resulting from the map $\mathbf{x} \mapsto P(\mathbf{x})$. The limit cycle is stable if all eigenvalues of this maps are inside the unit circle. (Taken from (Kuznetsov 1998))

---

[4] This is actually called asymptotic stability

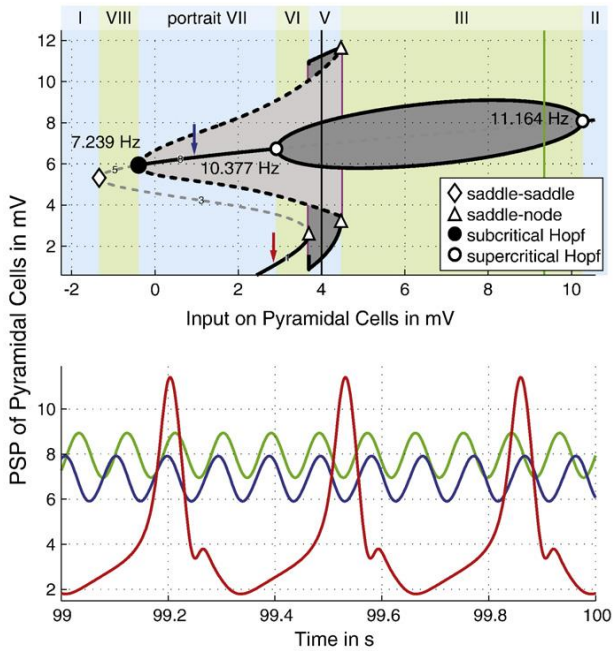

Figure 4. Bifurcation diagram and a few examples of time series of the J&R model with input only to the pyramidal population. The bifurcation diagram was divided in different regions corresponding to different phase portaits (not shown in this review available in (Spiegler 2010)). The blue (red) time series corresponds to region V with initial conditions at the blue (red) arrow. The green time series corresponds to the green line in region III. Several types of bifurcations are shown (see Kutnetsov 1998 for a more detailed explanation of all of them) (Taken from (Spiegler et al. 2010)).

points in the parameter space (see Figs. 3, 4 and 5).

## III. DISCUSSION OF EXAMPLES REVIEWED IN THE LITERATURE

We want to clarify that, since this only a project paper of a course, we shall deliberately take some figures from various published papers, copy them into the text and discuss them, without original the author's compliances.

### A. The Jansen & Rit Model

We shall start with the famous Jansen & Rit model (J&R) (Jansen and Rit 1995). This is a model of one single mass comprising populations E, e and i. Population e (spiny stellate in layer IV) is the only population receiving external inputs. Second order dynamics are ruled by equal decay and rise times $\tau_{pq}^r = \tau_{pq}^d = \tau_q$. Also there is no wave propagation, i.e. $\phi_{pq} = Q_q$ in (1.2).

The Fig. 4 shows a bifurcation diagram, obtained by (Spiegler et al. 2010), of the J&R and the corresponding time series for different parameter values. The bifurcation diagram shows the membrane potential of population E versus the average magnitude of a constant input to that population. Several types of bifurcations are depicted in Fig. 4. For example, a supercritical Andronov-Hopf bifurcation (Super-

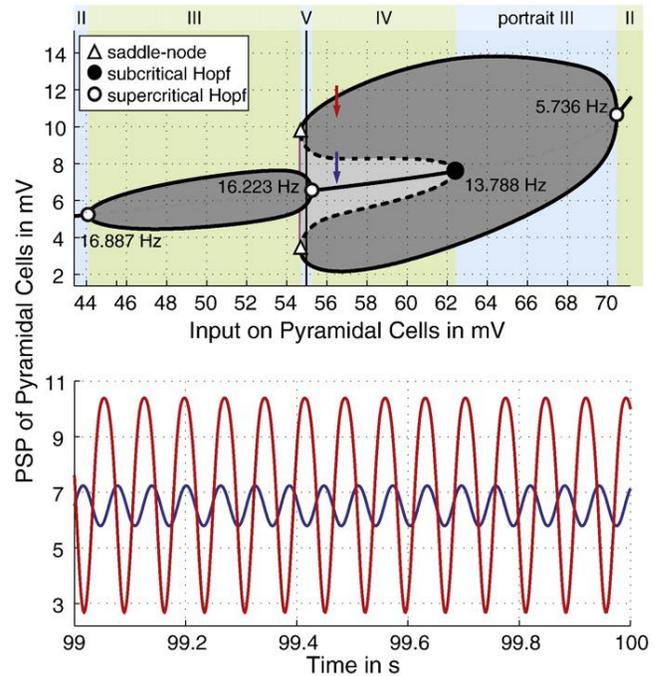

Figure 5. Bifurcation diagram and a few examples of time series of the J&R model with input to the three neuron populations and $v_e = -17\,\text{mV}$, $v_i = 4\,\text{mV}$, $\tau_e = 4\,\text{ms}$ and $\tau_i = 22\,\text{ms}$. The blue (red) time series corresponds to region IV with initial conditions in the blue (red) arrows. If the external input to the pyramidal cells is incresed beyond the subcritical AH bifurcation, time series will be all like the red one. Only by decreasing the input to region III an then back again to region V the system can recover the low amplitud oscillations represented by the blue time series (Taken from (Spiegler et al. 2010)).

AH1[5]) occurs. This means that for an increase of the input beyond a certain value ($\sqcup$ 3mV), a stable limit cycle appears around an unstable equilibrium which was stable before the bifurcation.

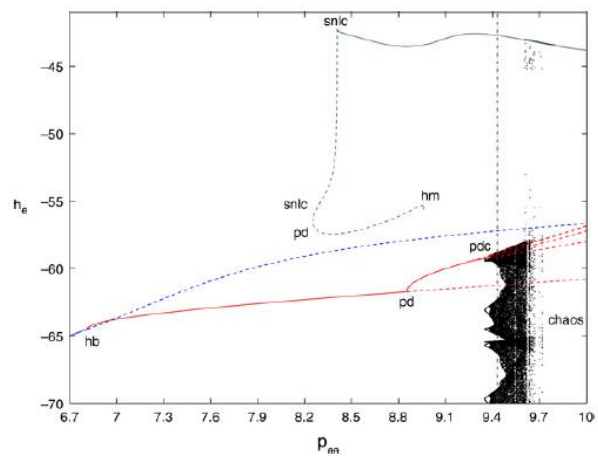

Figure 6. Bifurcation diagram of Liley's model (Liley). Apart from a rich repertoire of bifurcations, we want to note that chaos arises for inputs to the excitatory cells higher than 9.2 (Taken from (Dafilis et al. 2009))

---

[5] The "1" is to differentiate it from the rightmost AH bifurcation, which would be Super-AH2. We follow the same rule in Fig. 4 for subcritical AH bifurcations.

As clearly stated by the authors, if the initial condition is the blue arrow, and the input is increased beyond the Super-AH1, the system travels from the stable equilibrium to the basin of attraction of a stable limit cycle. This will change the time series from a constant (or stochastically driven damped oscillations of very low amplitude around a focus) to sinusoidal medium amplitudes (blue time series).

On the other hand, when starting from the red arrow, an increase of the input beyond the Super-AH1 drives the system to the basin of attraction of a quite higher amplitude anharmonic cycle (red time series). This is due to a global bifurcation named saddle-node bifurcation of the Poincaré map (Kuznetsov 1998). This high amplitude spike-like waveform might correspond to epileptic seizures.

In summary, we could say from the bifurcation diagram in Fig. 4 that a "smart choice" of initial conditions that would cope with a sudden increase of the input to pyramidal cells without generating epileptic seizures is to restrict the activity of pyramidal cells within the interval 4-8 mV. Note that in this interval the system remains within the outer basin of attraction of the limit cycle beyond the Super-AH1.

### B. Jansen & Ritt with inputs to all populations

Inputs to populations e and i were also included in (Spiegler et al. 2010). This changed drastically the phase portraits and bifurcations diagrams of the original J&R model. Fig. 5 shows the bifurcation diagram for $\upsilon_e = -17\,\text{mV}$, $\upsilon_i = 4\,\text{mV}$, $\tau_e = 4\,\text{ms}$ and $\tau_i = 22\,\text{ms}$.

Starting at the blue arrow, an increase of the input to the pyramidal cells beyond Sub-AH will put the system within the basin of attraction of a limit cycle with much higher amplitudes (red time series) than the ones during related to the focus (blue time series). This process is *irreversible* since a decrease of the same amount of input will keep the system dwelling in the limit cycle. It is actually necessary to decrease even more the input beyond Super-AH2 and return again to the focus basin to reach the initial regime. A smart strategy of the brain in this case would be to allow for flexible input values in order to overcome possible epileptic seizures

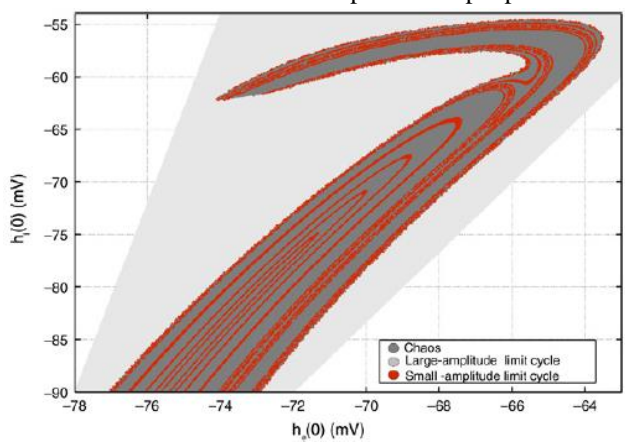

Figure 7. Extremely intricate patterns of dependency on the initial conditions in the Liley's model. The system can reach two stalble limit cycles and a chaotic attractor depending on the initial values of both neural populations ) (Taken from (Dafilis et al. 2009)).

provoked by excessive excitation of pyramidal cells.

### C. Dependency on initial conditions in Liley's model

In (Dafilis et al. 2009) Liley's model (Liley, Cadusch, and Dafilis 2002) is used to investigate the dependency of Lyapunov Exponents and the presence of chaos on a space of initial conditions. Liley's model is a bit different from the NNM and NFM presented in subsection A of section II. It incorporates voltage dependent synaptic conductances that are shunted around the so-called reversal potential. This slight modification has a drastic impact on the dynamical behavior of the system, its multistability and bifurcation diagrams (see Fig. 6, chaos is achieved for high inputs on excitatory neurons). It is shown in (Dafilis et al. 2009) that generalized multistability can occur in this model for the proper set of parameters, i.e. chaotic, small and high amplitude limit cycles coexist. Although no fractal evidence was found, the distribution of the different basins of attraction for these three attractors is very intricate, as it can be seen in Fig. 7. This demonstrates the high importance and dependency on the initial conditions. Since it is speculated that chaos in the brain (as well as marginal instability) might be related to a property of a system to be able to learn, a "smart choice of the brain" might be to keep a high cortical excitability for certain decision making and learning tasks.

### IV. CONCLUSIONS

In this short review we have shown different examples that evidence the high dependency on the initial conditions of neural models of the brain. We firstly have provided the basic ingredients of neural field models. We also provided the basic notions on the analysis of dynamical systems and its evolution equations which are necessary to understand terms as stability, generalized multistability and bifurcations. A comprehensive analysis of the time series corresponding to the orbits in the state space gave insights into the behavior of neural models in the way that it is usually presented by experimentalists.

We think that this could be a start to more detailed and systematic reviews (and further studies) on the effect of initial conditions on the behavior of mean field brain models.